\documentclass[letterpaper, 10 pt, conference]{ieeeconf}  

\IEEEoverridecommandlockouts                              

\usepackage{cite}
\usepackage{amsmath,amssymb,amsfonts}
\usepackage{graphicx}
\usepackage{textcomp}
\usepackage[table]{xcolor}
\usepackage{booktabs}
\usepackage{multirow}
\usepackage{makecell}
\usepackage{array}
\usepackage{hyperref}
\usepackage{adjustbox}

\title{ \textbf{Bidirectional Modeling of mmWave Human Signals for 
\\
Enhancement \& Synthesis}}

\author{
Shuntian Zheng, Hoang Hai Pham, Jiaqi Li, Xiaoman Lu, Shuai He, Yu GUAN*\\
shuntian.zheng,Hoang-Hai.Pham,Jiaqi.Li.16,Xiaoman.Lu@warwick.ac.uk,
hs19951021@bupt.edu.cn,\\
yu.guan@warwick.ac.uk
}

\begin{document}

\maketitle
\thispagestyle{empty}
\pagestyle{empty}

\begin{abstract}
Millimeter-wave (mmWave) is attractive for human-centered robotics in smoke, darkness, and privacy-sensitive indoor spaces: it measures reflections rather than visually identifiable imagery and remains usable without light.
Deployed systems, however, face two complementary needs that run in opposite directions.
Enhancement starts from a raw range–angle–Doppler (RAD) heatmap, where human returns coexist with nonhuman structure, and aims to isolate the human signal by removing nonhuman returns, yielding a clean RAD input for downstream perception.
Synthesis starts from denser human geometry and aims to synthesize natural RAD observations, including environmental and hardware features, thereby expanding scarce supervision data.
Although the two tasks reverse direction, both require modeling how human and nonhuman reflections coexist in the same measurement. Among RAD axes, Doppler provides motion information native to the modality, with the potential to support this modeling.
We therefore cast both tasks on the dual-domain RAD heatmap and propose HeatWave, which injects Doppler when it is locally reliable, and routes Doppler asymmetrically for enhancement and synthesis within one shared backbone.
This design improves both tasks: enhanced observations reduce mean per-joint position error by $11.4\%$ relative to the raw recording, and synthesized samples that replace half of the real radar labels recover most of the activity accuracy lost to the reduced budget. 
Relative to an unspecialized dual-task heatmap baseline, the enhancement improves motion-critical samples under metallic clutter by $2.3\times$ more than when geometry already suffices, and synthesis follows the same ordering with a narrower spread, providing the community with a research direction more closely aligned with the signal's characteristics.
\end{abstract}

\section{Introduction}

Human-aware robots need to estimate where people are and how they move under conditions that often defeat cameras: smoke, darkness, glare, and privacy-sensitive spaces such as homes, clinics, and factories~\cite{deng2023midas++}.
Millimeter-wave (mmWave) radar is a practical sensing front end for these settings because it measures electromagnetic reflections rather than appearance, remains usable without ambient light, and does not record visually identifiable imagery~\cite{yang2025mmdear}.
It offers a further advantage that vision sensors cannot match under their imaging physics, and that matters for robots reasoning about trajectories, gestures, or short-horizon forecasts: \textbf{a single measurement encodes instantaneous motion~\cite{zheng2026doppler}.}

\begin{figure}[t]
\centering
\includegraphics[width=\linewidth]{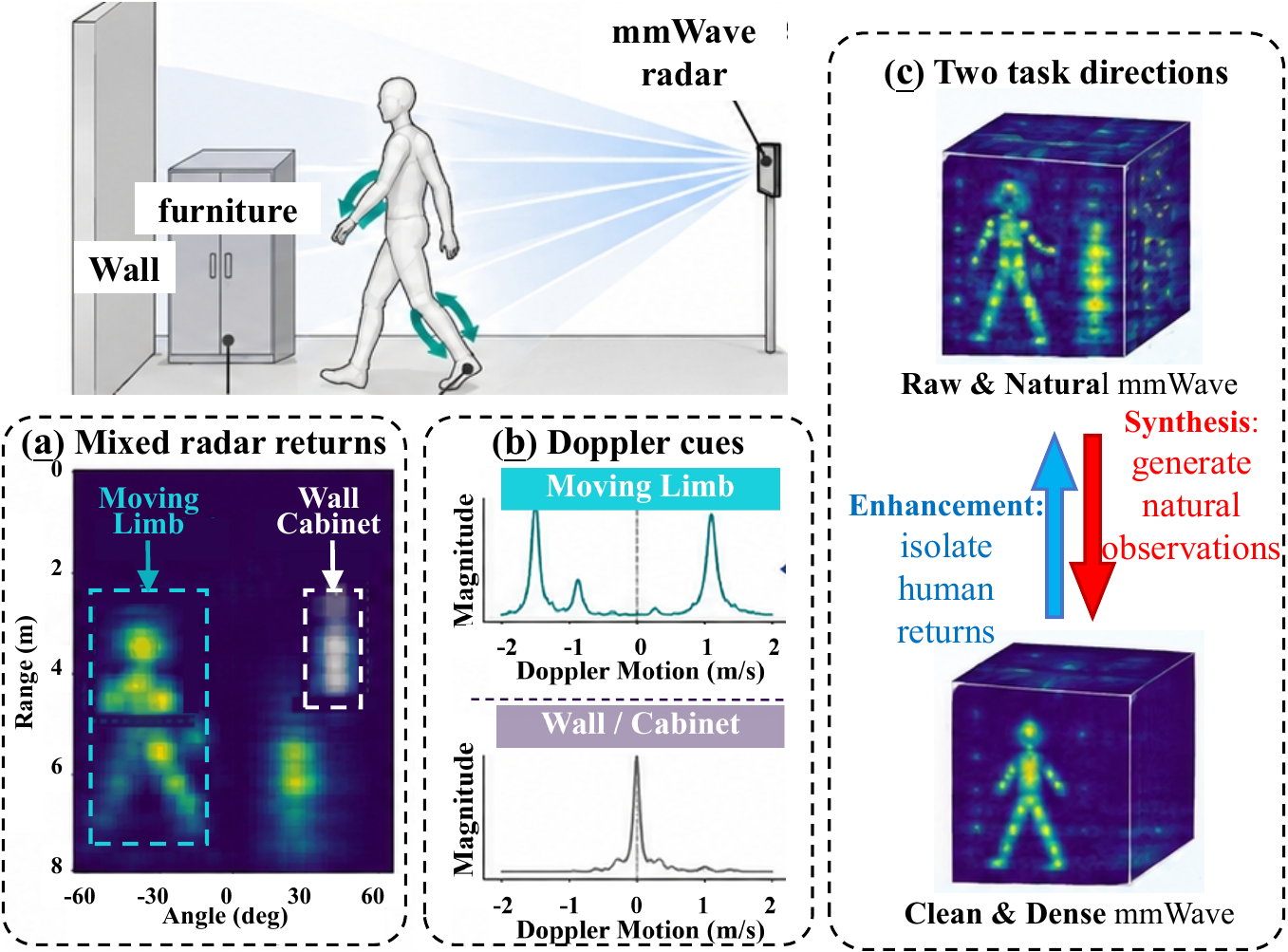}
\caption{mmWave mixes a walking person with static furniture. (a)~A limb and a cabinet can occupy similar spatial support, (b)~yet only the limb produces a multi-peak Doppler spectrum. (c)~Enhancement and synthesis are reverse flows.}

\label{fig:teaser}
\end{figure}

A single Frequency-Modulated Continuous Wave (FMCW) mmWave frame organizes human and non-human returns as a range--angle--Doppler (RAD) heatmap~\cite{iovescu2020fundamentals}.
Spatial magnitude on the range--angle grid supports body localization and coarse shape, while the Doppler axis reports radial velocity~\cite{fan2024enhancing}.
The two can disagree as shown in Fig.~\ref{fig:teaser}; a moving limb and a static cabinet may occupy similar spatial support, yet only the limb produces a multi-peak Doppler spectrum.
We refer to this magnitude-plus-Doppler structure as the \textbf{dual-domain RAD heatmap}, which separates mmWave from cameras and LiDAR~\cite{luan2024diffusion}.

In deployment, however, robots rarely observe clean, richly labeled RAD heatmaps: a raw recording mixes human returns with multipath from walls and furniture and with hardware-dependent artifacts, and large paired corpora for supervision are expensive to collect.
mmWave human sensing therefore needs both \textbf{enhancement}, which isolates the human-borne signal from a cluttered raw recording~\cite{zheng2026doppler}, and \textbf{synthesis}, which generates realistic recordings from human geometry to relieve label scarcity~\cite{deng2023midas++}.

Existing pipelines, however, rarely keep the dual-domain structure intact.
Enhancement methods built on Constant False Alarm Rate (CFAR) detection and clustering super-resolve sparse points or LiDAR-like targets after Doppler and much of the spatial spectrum have already been thinned away~\cite{fan2024diffusion}, while synthesis methods optimize fidelity under strong geometric assumptions or scene-specific simulators~\cite{deng2023midas}.
Heatmap-based pose estimators show that RAD inputs carry richer human information than points alone~\cite{lee2023hupr,choi2025mvdoppler}, yet assume the observation is already given. Concurrent unified models do couple the two tasks, since both need to account for how human and non-human returns co-exist, but they operate on point sets, so mmWave is still handled as sparse geometry.
For robots interpreting dynamic humans, these choices discard motion evidence that geometry alone often cannot replace.
The bottleneck, we argue, is not model capacity but the problem setting itself: \textbf{enhancement and synthesis should be posed on the dual-domain RAD heatmap, with Doppler remaining available for reasoning}.

We attribute this bottleneck to three mismatches.
\textbf{(1) Premature geometric collapse} converts RAD heatmaps into unordered points before learning, deleting Doppler channels and spatial spectrum that later modules often cannot recover.
\textbf{(2) Role-symmetric Doppler fusion} encodes magnitude and Doppler as interchangeable channels, ignoring that Doppler is motion-sensitive but not human-exclusive: clutter, multipath, and hardware effects inject spurious spectral activity that can destabilize predictions if trusted blindly~\cite{zheng2026learn}.
\textbf{(3) Task-agnostic Doppler use} arises once the two tasks share one backbone that preserves Doppler: enhancement needs to read observed spectra to separate human returns from non-human clutter, whereas synthesis needs to invent motion-consistent spectra and plausible non-human structure from geometry.

We therefore formulate unified enhancement and synthesis as \textbf{Doppler-preserving dual-task modeling} and instantiate it with \textbf{HeatWave}, whose three modules address one mismatch each: it encodes magnitude and Doppler on a shared range--angle grid so that the dual-domain structure survives; injects Doppler only where it is locally reliable, so motion cues guide structure without symmetrically overwriting it; and routes that cue by task, so the two directions share one backbone without sharing one Doppler role.
The formulation matters most in \textbf{motion-critical} samples, where sparse geometry is ambiguous across actions or limb velocities while Doppler remains informative.

Across three radars, HeatWave improves both tasks over point-cloud and heatmap baselines.
Enhanced observations reduce mean per-joint position error by \textcolor{black}{$11.4\%$} relative to the raw recording and by \textcolor{black}{$4.7\%$} relative to an unspecialized heatmap control that already couples the two directions on a shared backbone but leaves Doppler unspecialized.
The remaining improvement over that control is \textcolor{black}{$2.3\times$} larger under metallic clutter for enhancement than when geometry already suffices, and \textcolor{black}{$1.4\times$} for synthesis.

The main contributions of this work are:

\textbf{C1. RAD-level problem setting for robot sensing.}
We cast unified mmWave enhancement and synthesis as dual-task modeling, identifying geometric collapse, role-symmetric Doppler fusion, and task-agnostic Doppler use as obstacles to exploiting mmWave's motion cues.

\textbf{C2. HeatWave: complementary Doppler-centric modules.}
We adapt confidence-gated Doppler conditioning into a generative dual-domain decoder and introduce task-conditioned routing with a synthesis-side Doppler prior, together mapping one-to-one onto the three mismatches above.

\textbf{C3. Motion-critical evidence and downstream transfer.}
Through per-device evaluation on three radars and a motion-critical stratification of both task directions, we show that HeatWave's gains concentrate where geometry-only pipelines degrade, benefiting downstream human perception metrics relevant to indoor robots.

\section{Related Work}


\subsection{mmWave enhancement for robot perception}
Robotics systems densify or denoise radar for mapping, odometry, and human reconstruction when cameras fail, using CFAR clustering or learning-based super-resolution~\cite{ren2024tiger}.
These methods improve geometric completeness, but they typically supervise on sparse points or top-down maps and discard Doppler channels before learning.
That practice is premature geometric collapse: later modules cannot recover the spectra that separate human returns from clutter and multipath~\cite{chen2006micro}.

\subsection{mmWave synthesis for data expansion}
Physics-based and learning-based generators synthesize Doppler signatures, spectrograms, or radar returns from vision, meshes, or scene parameters, sometimes with explicit propagation priors~\cite{deng2023midas}.
They reduce labeling cost, yet they optimize a synthesis-only objective and do not learn the reverse enhancement direction on the same dual-domain RAD heatmap, so task-agnostic Doppler use remains unresolved for the joint setting studied here.

\subsection{Heatmap sensing and unified enhancement--synthesis}
Heatmap-based human sensing models take range--angle or RAD heatmaps as input and report richer pose and activity cues than sparse points~\cite{lee2023hupr,choi2025mvdoppler}, yet they typically assume a usable observation is already given.
When they use magnitude and Doppler together, they often encode them as interchangeable channels, which is role-symmetric Doppler fusion~\cite{lee2023hupr,choi2025mvdoppler}.
Methods that retain RAD heatmaps similarly preserve the motion axis~\cite{zheng2026doppler}, but target discriminative perception or driving-scene rendering rather than coupled enhancement and synthesis; Sec.~\ref{sec:method} adapts the confidence-gated Doppler mechanism of the former into this generative dual-task setting through task-conditioned routing.
mmWaveFlow unifies the two directions through invertible transport on human mmWave point clouds~\cite{su2026mmwaveflow}.
Both tasks need to model how human and non-human returns co-exist; we retain it on the dual-domain RAD heatmap so that Doppler can take the task-dependent roles that point cloud sets can't supports.

\section{Preliminaries and Problem Setup}
\label{sec:setup}

The two directions reviewed above are usually posed after Doppler has been discarded.
We now define the representation, the two tasks, and the requirements they induce.

\noindent\textbf{Dual-domain RAD heatmaps.}
An FMCW radar forms, at time $t$, a non-negative range--angle--Doppler (RAD) magnitude heatmap $\mathbf{H}_t\in\mathbb{R}^{R\times A\times D}$ after standard FFTs along the three axes~\cite{iovescu2020fundamentals,lee2023hupr}, where $R$, $A$, and $D$ index range, angle, and Doppler bins.
To keep the two roles named in Sec.~I explicit, we read $\mathbf{H}_t$ as a spatial-magnitude map and a Doppler-spectrum field on one range--angle grid,
\begin{equation}
\small
\mathbf{v}_{t,r,a}=\mathbf{H}_t[r,a,:]\in\mathbb{R}^{D},
\quad
\mathbf{S}_t[r,a]=\frac{1}{D}\sum_{b=1}^{D}\mathbf{H}_t[r,a,b],
\label{eq:dual_domain}
\end{equation}
writing $\mathbf{V}_t=\{\mathbf{v}_{t,r,a}\}$ for the spectrum field, which re-indexes $\mathbf{H}_t$ rather than transforming it.
$\mathbf{S}_t$ carries spatial structure for localization, while $\mathbf{V}_t$ carries the motion-sensitive spectra that peak extraction to points discards.

\noindent\textbf{RAD-level enhancement and synthesis.}
Let $\mathbf{H}$ denote a raw recording, which mixes human returns with static environmental reflections and hardware-dependent perturbations, and let $\mathbf{H}_{\mathrm{h}}$ denote a human-only reference on the same grid, retaining energy inside the human support.
Let $X_{\mathrm{dense}}$ denote a dense human geometric condition available at training time, such as a pose-aligned surface from vision-, LiDAR-, or motion-capture-based annotation.
\emph{Enhancement} seeks $\hat{\mathbf{H}}\approx\mathbf{H}_{\mathrm{h}}$ from $\mathbf{H}$, that is, separation of the human-borne signal from non-human returns rather than generic denoising; \emph{synthesis} seeks $\hat{\mathbf{H}}\approx\mathbf{H}$ from $X_{\mathrm{dense}}$, that is, generation of a realistic recording with plausible non-human structure and motion-consistent Doppler, so that scarce labeled radar can be expanded.
We treat the two as one dual-task family with direction $d\in\{\mathrm{enh},\mathrm{gen}\}$ and task target $\mathbf{H}^{\star}$, equal to $\mathbf{H}_{\mathrm{h}}$ for enhancement and to $\mathbf{H}$ for synthesis:
\begin{equation}
\small
\hat{\mathbf{H}}=F_{\theta}(X_{\mathrm{in}};d),\qquad
X_{\mathrm{in}}=\begin{cases}
\mathbf{H}, & d=\mathrm{enh},\\
X_{\mathrm{dense}}, & d=\mathrm{gen}.
\end{cases}
\label{eq:dual_task}
\end{equation}
The shared family $F_{\theta}$ thus outputs a RAD heatmap in both directions but moves toward mirrored targets; keeping both on the grid of~\eqref{eq:dual_domain}, rather than translating among point sets, is what lets Doppler take task-dependent roles.

\noindent\textbf{Motion-critical regimes.}
Geometry alone can suffice when a person is nearly static, whereas indoor robots depend on motion cues most during walking or close-range interaction.
To separate these, we score the dynamics of a sample $x$ from its ground-truth joint positions $\{\mathbf{p}_j\}_{j=1}^{J}$ and velocities $\dot{\mathbf{p}}_j$:
\begin{equation}
\small
\mathrm{MCS}(x)=\mathrm{PercentileRank}\!\left(\frac{1}{J}\sum_{j=1}^{J}\lVert\dot{\mathbf{p}}_j\rVert_2\right),
\label{eq:mcs}
\end{equation}
where $\mathrm{PercentileRank}$ maps the mean joint speed to $[0,1]$ within the evaluation set, and a threshold $\tau$ yields a geometry-sufficient regime with $\mathrm{MCS}<\tau$ and a motion-critical regime with $\mathrm{MCS}\ge\tau$.
MCS is a pose-based proxy for dynamic difficulty; it marks samples where sparse geometry is more likely to be ambiguous.

\noindent\textbf{Requirements.}
This setup induces three requirements that organize Sec.~\ref{sec:method}, each countering one mismatch of Sec.~I:
R1, \emph{Preserve-RAD}, keep modeling on the dual-domain grid of~\eqref{eq:dual_domain}, against premature geometric collapse;
R2, \emph{Asymmetric-Doppler}, weight Doppler by how reliable the local spectrum is and let it act as a one-way motion cue rather than as a channel interchangeable with spatial magnitude, against role-symmetric Doppler fusion;
R3, \emph{Task-conditioned Doppler routing}, assign different Doppler roles to the two directions inside~\eqref{eq:dual_task}, against task-agnostic Doppler use.
How much reliability weighting a direction needs depends on whether its spectrum is observed or predicted, which is one reason the roles in R3 cannot be shared.


\begin{figure*}[t]
\centering
\includegraphics[width=\linewidth]{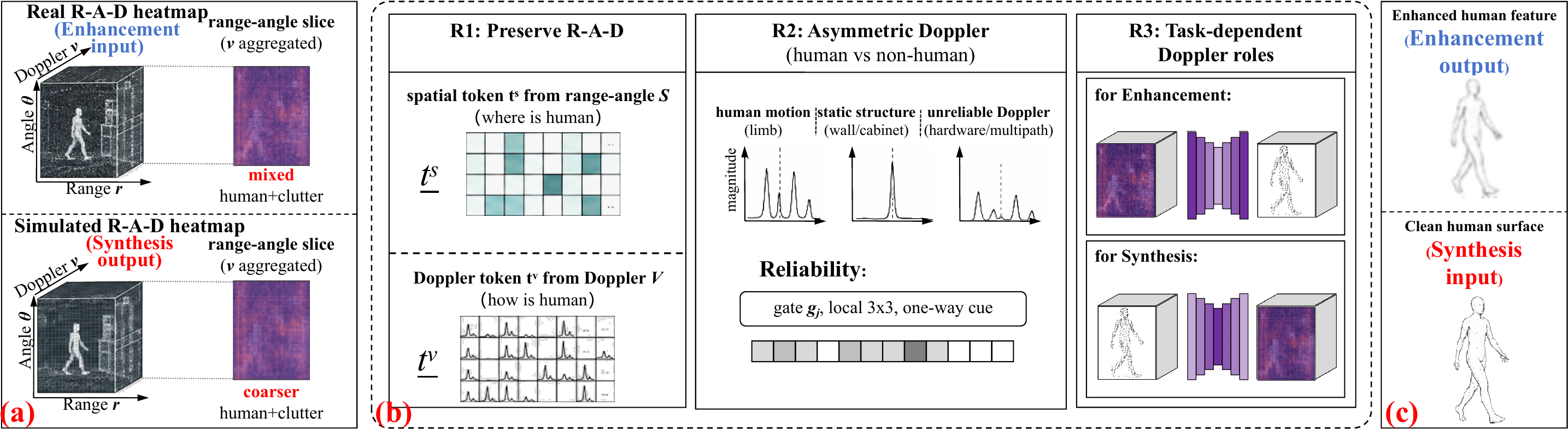}
\caption{Workflow of HeatWave. (a)~Mixed RAD is the enhancement input and synthesis target. (b)~The shared core keeps dual-domain tokens (R1), gates Doppler by local reliability (R2), and assigns opposite Doppler roles (R3). (c)~Enhancement returns human features, while synthesis is conditioned on this.}
\label{fig:method}
\end{figure*}

\section{Method}
\label{sec:method}


\subsection{Dual-Domain RAD Encoder}
Premature geometric collapse deletes Doppler bins before learning, and no later module can restore the motion evidence they carry.
To satisfy R1, the encoder stays on the RAD grid and builds two token streams, that is, two sets of embedded vectors indexed by the same range--angle geometry.

Given the RAD input $\mathbf{H}$, we form $\mathbf{S}$ and $\mathbf{V}$ as in~\eqref{eq:dual_domain}, dropping the frame index $t$ because all operations below act within one frame.
Spatial tokens come from non-overlapping $p\times p$ patches of $\mathbf{S}$ with $p$=4, while Doppler tokens remain one per cell so motion is not pooled away before conditioning:
\begin{equation}
\small
\mathbf{t}^{s}_{i}=f_{s}(\mathbf{S}^{(i)})\in\mathbb{R}^{C},
\qquad
\mathbf{t}^{v}_{j}=f_{v}(\mathbf{v}_{r,a})\in\mathbb{R}^{C},
\label{eq:tokens}
\end{equation}
where $f_{s}$ is a convolutional patch encoder, $f_{v}$ is an MLP over the $D$-bin spectrum at cell $(r,a)$, $i$ indexes a patch, $j$ indexes a cell, and $C$ is the token width.
This dual-stream split follows the tokenization used by PULSE for Doppler-conditioned pose prompting~\cite{zheng2026doppler}; here the tokens instead feed the generative dual-domain decoding of~\eqref{eq:decode}, so every token must remain sufficient to reconstruct the full RAD grid rather than a compact pose vector.
For synthesis, a point-set encoder $E_{\mathrm{dense}}$ maps $X_{\mathrm{dense}}$ to a global latent $z_{\mathrm{dense}}=E_{\mathrm{dense}}(X_{\mathrm{dense}})\in\mathbb{R}^{C}$, which provides the source state when $d=\mathrm{gen}$.
The target $\mathbf{H}^{\star}$ is decomposed by the same operators, so prediction and reference are always compared on the dual-domain grid.
Because structure and motion are carried on one grid rather than in an unordered point set, every later stage still has Doppler available.

\subsection{Reliability-Gated Doppler Conditioning}
Doppler is motion-sensitive but not human-exclusive: clutter, multipath, and hardware can inject spectral activity that resembles limb motion~\cite{chen2006micro}.
Role-symmetric Doppler fusion therefore risks writing nuisance velocities into the structural pathway.
To satisfy R2, HeatWave treats Doppler as a one-way condition: motion cues may modulate spatial tokens, but Doppler is not encoded as an interchangeable parallel channel.
The gating primitive builds on the confidence-gated cross-attention~\cite{zheng2026doppler}; HeatWave conditions a generative decoder that reconstruct a full RAD in both magnitude and spectral shape, and embeds the gate inside the task-conditioned routing, which pairs it with a synthesis-side Doppler prior.
Each Doppler token receives a gate:
\begin{equation}
\small
g_{j}=\sigma\!\big(f_{g}(\mathbf{t}^{v}_{j})\big)\in[0,1],
\label{eq:gate}
\end{equation}
where $\sigma$ is the logistic function and $f_{g}$ is a small MLP, so $g_{j}$ is a learned score indicating how far the local spectrum should influence spatial estimation.
Spatial token $\mathbf{t}^{s}_{i}$ then attends only inside a local neighborhood $\mathcal{N}(i)$ on the shared grid, taken as the cell-level Doppler tokens whose $(r,a)$ fall inside the 3$\times$3 block of patches (that is, 12$\times$12 cells) centered on the patch of $\mathbf{t}^{s}_{i}$, so the neighborhood covers that patch's own footprint together with surrounding context, with gate-biased weights
\begin{equation}
\small
\alpha_{ij}=\frac{\exp\!\Big(\frac{(\mathbf{t}^{s}_{i}W_{Q})(\mathbf{t}^{v}_{j}W_{K})^{\top}}{\sqrt{d_{k}}}+\beta g_{j}\Big)}{\sum_{j'\in\mathcal{N}(i)}\exp\!\Big(\frac{(\mathbf{t}^{s}_{i}W_{Q})(\mathbf{t}^{v}_{j'}W_{K})^{\top}}{\sqrt{d_{k}}}+\beta g_{j'}\Big)},
\label{eq:gated_attn}
\end{equation}
where $d_{k}$ is the key dimension and $\beta$ controls the strength of the gate bias.
The resulting motion context $\mathbf{c}_{i}=\sum_{j\in\mathcal{N}(i)}\alpha_{ij}(\mathbf{t}^{v}_{j}W_{V})$ updates the spatial token by
\begin{equation}
\small
\tilde{\mathbf{t}}^{s}_{i}=\mathbf{t}^{s}_{i}+\lambda_{i}\mathbf{c}_{i},\qquad \lambda_{i}=\sigma\!\big(f_{\lambda}(\mathbf{t}^{s}_{i})\big).
\label{eq:spatial_update}
\end{equation}
Locality limits cross-region spectral leakage in metal-rich indoor scenes, while the gate down-weights spectra that are active but pose-irrelevant.
The update in~\eqref{eq:spatial_update} is therefore one-way by construction: Doppler can constrain an ambiguous body configuration, but it never overwrites the spatial stream on which localization depends.

\subsection{Task-Conditioned Doppler Routing}
\label{sec:task_routing}
The two directions use Doppler in opposite ways.
Under enhancement, Doppler is present in the raw recording, so the model uses it to preserve human-borne spectra and down-weight non-human returns, which concentrate near zero Doppler.
Under synthesis, Doppler is absent from $X_{\mathrm{dense}}$, the model instead predicts motion-consistent spectra, together with plausible non-human structure, and the output resembles a real recording rather than a geometry-only heatmap.

To satisfy R3, HeatWave introduces a task embedding and routes gated Doppler context accordingly.
Let
\begin{equation}
\small
\mathbf{e}_{d}=\mathrm{Embed}(d)\in\mathbb{R}^{C},\qquad d\in\{\mathrm{enh},\mathrm{gen}\}.
\label{eq:task_embed}
\end{equation}
The gate and residual scale become task-dependent,
\begin{equation}
\small
g_{j}^{(d)}=\sigma\!\big(f_{g}([\mathbf{t}^{v}_{j};\mathbf{e}_{d}])\big),\qquad
\lambda_{i}^{(d)}=\sigma\!\big(f_{\lambda}([\mathbf{t}^{s}_{i};\mathbf{e}_{d}])\big),
\label{eq:task_gate}
\end{equation}
so~\eqref{eq:gated_attn}--\eqref{eq:spatial_update} are evaluated with $g_{j}^{(d)}$ and $\lambda_{i}^{(d)}$ in place of $g_{j}$ and $\lambda_{i}$.
What the gates weigh then differs by direction: under enhancement they discount observed spectra that clutter or multipath may have contaminated, whereas under synthesis they discount spectra the model has itself predicted, so reliability weighting need not matter equally in the two directions.
Because synthesis has no observed spectrum to gate, a lightweight prior head predicts a Doppler field from geometry.
A global latent alone cannot place spectra on the body, so $f_{\mathrm{prior}}$ is evaluated per cell,
\begin{equation}
\small
\mathbf{V}_{\mathrm{prior}}[r,a,:]=f_{\mathrm{prior}}\big([z_{\mathrm{dense}};\mathbf{g}_{r,a};o_{r,a}]\big),
\label{eq:doppler_prior}
\end{equation}
where $\mathbf{g}_{r,a}$ is the range--angle coordinate of the cell and $o_{r,a}$ is the rasterized occupancy of $X_{\mathrm{dense}}$ there.
The two-layer MLP therefore yields a spatially varying field in $\mathbb{R}^{R\times A\times D}$, which is encoded into Doppler tokens and routed by the same gated pathway, whereas enhancement draws those tokens from the observed recording $\mathbf{H}$.

\noindent\textbf{Dual-domain decoding.}
Because~\eqref{eq:spatial_update} updates only the spatial stream, the routed latent retains both streams, $z=(\{\tilde{\mathbf{t}}^{s}_{i}\},\{\mathbf{t}^{v}_{j}\})$, and the decoder reads each in its own role.
A transposed-convolutional branch $D^{s}_{\psi}$ returns the spatial tokens to the range--angle grid as per-cell features $\mathbf{u}_{r,a}$, and a spectral head $D^{v}_{\psi}$ forms the $D$ bins of that cell from $\mathbf{u}_{r,a}$ and the Doppler token of the same cell, written $\mathbf{t}^{v}_{r,a}$:
\begin{equation}
\small
\mathbf{u}_{r,a}=D^{s}_{\psi}\big(\{\tilde{\mathbf{t}}^{s}_{i}\},\mathbf{e}_{d}\big)[r,a],
\quad
\hat{\mathbf{H}}[r,a,:]=D^{v}_{\psi}\big(\mathbf{u}_{r,a},\mathbf{t}^{v}_{r,a},\mathbf{e}_{d}\big).
\label{eq:decode}
\end{equation}
The spatial branch therefore decides where return energy is placed, while the spectral head decides how that energy is distributed over radial velocity, which is the quantity the spectral term of the objective below supervises.
One backbone thus serves both directions while Doppler keeps a task-dependent role, rather than becoming an interchangeable channel or an early-collapsed geometric feature.

\subsection{Training Objective}
Both directions are supervised on the dual-domain grid, so the reconstruction term compares prediction and target in each domain of~\eqref{eq:dual_domain},
\begin{equation}
\small
\mathcal{L}_{\mathrm{rec}}=\big\|\hat{\mathbf{H}}-\mathbf{H}^{\star}\big\|_{1}+\gamma\big\|\hat{\mathbf{S}}-\mathbf{S}^{\star}\big\|_{1},
\label{eq:rec}
\end{equation}
where $\hat{\mathbf{S}}$ is obtained from $\hat{\mathbf{H}}$ by the spatial aggregation in~\eqref{eq:dual_domain}, and $\mathbf{S}^{\star}$ is the corresponding map of $\mathbf{H}^{\star}$.
Since~\eqref{eq:rec} already supervises the magnitude of every RAD cell, the Doppler term constrains the shape of each spectrum instead.
Writing $\bar{\mathbf{v}}_{r,a}=\mathbf{v}_{r,a}/(\lVert\mathbf{v}_{r,a}\rVert_{1}+\epsilon)$ for a per-cell spectrum normalized to unit mass, with a small $\epsilon$ for stability, and taking $\bar{\mathbf{v}}_{r,a}$ and $\bar{\mathbf{v}}^{\star}_{r,a}$ from the prediction and the target,
\begin{equation}
\small
\begin{aligned}
\mathcal{L}_{\mathrm{dop}}&=\frac{1}{RA}\sum_{r,a}\big\lVert\bar{\mathbf{v}}_{r,a}-\bar{\mathbf{v}}^{\star}_{r,a}\big\rVert_{1},\\
\mathcal{L}&=\lambda_{\mathrm{rec}}\mathcal{L}_{\mathrm{rec}}+\lambda_{\mathrm{dop}}\mathcal{L}_{\mathrm{dop}}.
\end{aligned}
\label{eq:total}
\end{equation}
The term thus supervises which radial velocities occupy a cell, independently of how strong that cell's return is.
The gate in~\eqref{eq:gate} is learned from these two terms alone; no auxiliary gate or sparsity regularizer is used. Optimizer settings and loss weights are reported in Sec.~\ref{sec:exp}.

\begin{table*}[t]
\caption{Main dual-task heatmap results. Lower values are better. D2M denotes synthesis and M2D denotes enhancement.}
\label{tab:main}
\centering
\begin{adjustbox}{width=\linewidth,center}
\setlength{\tabcolsep}{4pt}
\renewcommand{\arraystretch}{1.12}
\begin{tabular}{l|cccccccccccc}
\Xhline{1.2pt}
\toprule
& \multicolumn{4}{c}{\textbf{HuPR}} & \multicolumn{4}{c}{\textbf{XRF55}} & \multicolumn{4}{c}{\textbf{mmRadPose}} \\
\cmidrule(lr){2-5}\cmidrule(lr){6-9}\cmidrule(lr){10-13}
\textbf{Model} & \multicolumn{2}{c}{\textbf{D2M}} & \multicolumn{2}{c}{\textbf{M2D}} & \multicolumn{2}{c}{\textbf{D2M}} & \multicolumn{2}{c}{\textbf{M2D}} & \multicolumn{2}{c}{\textbf{D2M}} & \multicolumn{2}{c}{\textbf{M2D}} \\
& \textbf{H-MAE} & \textbf{S-MAE} & \textbf{H-MAE} & \textbf{S-MAE} & \textbf{H-MAE} & \textbf{S-MAE} & \textbf{H-MAE} & \textbf{S-MAE} & \textbf{H-MAE} & \textbf{S-MAE} & \textbf{H-MAE} & \textbf{S-MAE} \\
\midrule
Tiger~\cite{ren2024tiger} & \textcolor{black}{0.179} & \textcolor{black}{0.147} & \textcolor{black}{0.167} & \textcolor{black}{0.137} & \textcolor{black}{0.204} & \textcolor{black}{0.169} & \textcolor{black}{0.192} & \textcolor{black}{0.158} & \textcolor{black}{0.197} & \textcolor{black}{0.165} & \textcolor{black}{0.182} & \textcolor{black}{0.151} \\
\rowcolor[HTML]{EFEFEF}
mmWaveFlow~\cite{su2026mmwaveflow} & \textcolor{black}{0.164} & \textcolor{black}{0.134} & \textcolor{black}{0.152} & \textcolor{black}{0.125} & \textcolor{black}{0.187} & \textcolor{black}{0.155} & \textcolor{black}{0.175} & \textcolor{black}{0.145} & \textcolor{black}{0.180} & \textcolor{black}{0.151} & \textcolor{black}{0.166} & \textcolor{black}{0.138} \\
HuPRModel~\cite{lee2023hupr} & \textcolor{black}{0.148} & \textcolor{black}{0.122} & \textcolor{black}{0.138} & \textcolor{black}{0.114} & \textcolor{black}{0.169} & \textcolor{black}{0.140} & \textcolor{black}{0.159} & \textcolor{black}{0.131} & \textcolor{black}{0.164} & \textcolor{black}{0.137} & \textcolor{black}{0.151} & \textcolor{black}{0.125} \\
\rowcolor[HTML]{EFEFEF}
PULSE~\cite{zheng2026doppler} & \textcolor{black}{0.134} & \textcolor{black}{0.110} & \textcolor{black}{0.125} & \textcolor{black}{0.103} & \textcolor{black}{0.153} & \textcolor{black}{0.127} & \textcolor{black}{0.144} & \textcolor{black}{0.119} & \textcolor{black}{0.148} & \textcolor{black}{0.124} & \textcolor{black}{0.137} & \textcolor{black}{0.113} \\
mmWaveFlow-H~\cite{su2026mmwaveflow} & \textcolor{black}{0.128} & \textcolor{black}{0.105} & \textcolor{black}{0.119} & \textcolor{black}{0.098} & \textcolor{black}{0.146} & \textcolor{black}{0.121} & \textcolor{black}{0.137} & \textcolor{black}{0.113} & \textcolor{black}{0.141} & \textcolor{black}{0.118} & \textcolor{black}{0.130} & \textcolor{black}{0.108} \\
HeatWave & \textbf{\textcolor{black}{0.101}} & \textbf{\textcolor{black}{0.082}} & \textbf{\textcolor{black}{0.094}} & \textbf{\textcolor{black}{0.076}} & \textbf{\textcolor{black}{0.118}} & \textbf{\textcolor{black}{0.097}} & \textbf{\textcolor{black}{0.109}} & \textbf{\textcolor{black}{0.089}} & \textbf{\textcolor{black}{0.112}} & \textbf{\textcolor{black}{0.093}} & \textbf{\textcolor{black}{0.103}} & \textbf{\textcolor{black}{0.084}} \\
\bottomrule
\Xhline{1.2pt}
\end{tabular}
\end{adjustbox}
\end{table*}

\section{Experiments}
\label{sec:exp}


\subsection{Experimental Setup}
\label{sec:exp_setup}

\subsubsection{Datasets.}
We evaluate on three public mmWave human benchmarks that expose per-frame Doppler and cover three different commercial radars: \textbf{HuPR}~\cite{lee2023hupr}, a controlled single-person indoor setting; \textbf{XRF55}~\cite{wang2024xrf55}, multi-person interactions across several rooms; and \textbf{mmRadPose}~\cite{engel2025advanced}, single-person motion under strong metallic reflections with high-precision OptiTrack pose labels; all heatmaps are resampled to $R\times A\times D=\textcolor{black}{64\times 64\times 16}$.
HuPR and mmRadPose release the RAD cube directly as $\mathbf{H}$; XRF55 releases only range--angle and range--Doppler maps, so we reconstruct $\mathbf{H}$ under one fixed recipe, shared by every heatmap method, that shares one Doppler spectrum shape across angles per range bin, so angle-resolved separation is examined only on the two datasets whose released cubes retain it.

\subsubsection{RAD targets.}
Enhancement and synthesis targets follow Sec.~\ref{sec:setup} directly, with $\mathbf{H}_{\mathrm{h}}$ built from the projected skeleton's human support at training time only, so enhancement at inference consumes $\mathbf{H}$ alone with no pose label, and $X_{\mathrm{dense}}$ a $\le\textcolor{black}{512}$-point skeleton sample supplied identically to every heatmap method.
A fixed cell-averaging CFAR detector builds the point clouds used only by the point-cloud baselines and the export of Sec.~\ref{sec:pc_export}, never our RAD input; splits follow official or published protocols.

\subsubsection{Implementation.}
All models are trained with AdamW ($2\times10^{-4}$ learning rate, cosine decay, batch size \textcolor{black}{$8$}, \textcolor{black}{$200$} epochs); HeatWave uses token width $C=\textcolor{black}{128}$, patch size $p=\textcolor{black}{4}$, loss weights $\lambda_{\mathrm{rec}}=\textcolor{black}{1.0}$, $\lambda_{\mathrm{dop}}=\textcolor{black}{0.5}$, $\gamma=\textcolor{black}{0.5}$, gate-bias strength $\beta=\textcolor{black}{1.0}$, and a two-layer MLP for $f_{\mathrm{prior}}$ in~\eqref{eq:doppler_prior}, with these values and the $3\times3$ neighborhood $\mathcal{N}(i)$ fixed across every dataset and direction under a single run per entry with a fixed seed.

\subsubsection{Tasks and metrics.}
Following~\eqref{eq:dual_task}, we write Dense-to-mmWave (D2M) for synthesis and mmWave-to-Dense (M2D) for enhancement, and report H-MAE/S-MAE as mean absolute error against the RAD target and its spatial map $\mathbf{S}$ from~\eqref{eq:dual_domain} after each cube is scaled to unit peak magnitude, plus Chamfer Distance (CD, in centimeters) after peak export since $L_{1}$ fidelity alone does not penalize over-smoothing.
Downstream we report MPJPE/MPJVE for a fixed pose estimator retrained on each observation variant and top-1 activity accuracy when synthesized samples augment scarce real labels; motion criticality follows~\eqref{eq:mcs} at $\tau=\textcolor{black}{0.67}$, the upper tercile of the evaluation set.

\begin{figure*}[t]
\centering
\includegraphics[width=\linewidth]{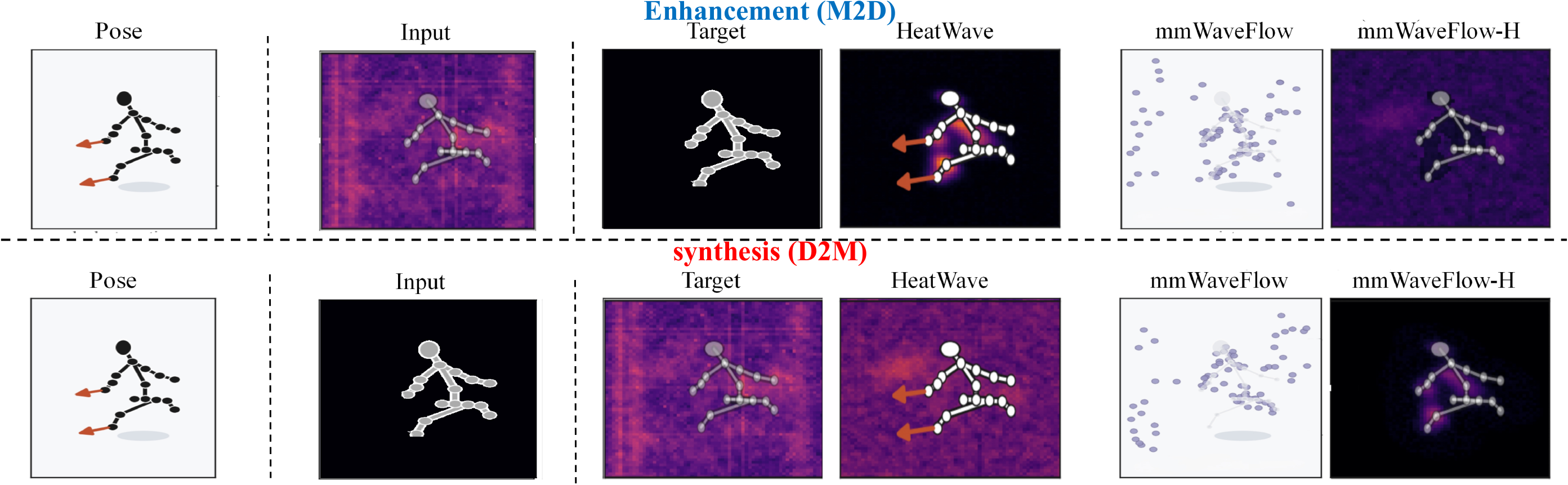}
\caption{Qualitative dual-task comparison. Pose and arrows mark body coordinates and the joints with the largest motion. mmWaveFlow emits points with no Doppler grid. Enhancement maps mixed $\mathbf{H}$ to human-only $\mathbf{H}_{\mathrm{h}}$; synthesis maps $X_{\mathrm{dense}}$ to mixed $\mathbf{H}$. HeatWave keeps body-occupied cells that mmWaveFlow-H suppresses; limb Doppler follows the moving joints and remains locally distinct from the target.}
\label{fig:qual_main}
\end{figure*}

\subsubsection{Baselines.}
\textbf{Tiger}~\cite{ren2024tiger}, a generic point-cloud diffusion model, and \textbf{mmWaveFlow}~\cite{su2026mmwaveflow}, to our knowledge the only concurrent method coupling enhancement and synthesis in one model, are serve as point-cloud-native references, operating entirely on collapsed point clouds rather than the dual-domain grid.
\textbf{PULSE}~\cite{zheng2026doppler} and \textbf{HuPRModel}~\cite{lee2023hupr} instead keep their published magnitude--Doppler encoders on the RAD lattice; because both were designed to regress human pose rather than a RAD heatmap, we retrain each with our dual-task heatmap decoder in place of its pose-regression head, isolating their encoder design from ours under an identical enhancement/synthesis objective while leaving PULSE's reliability-gated prompting and HuPRModel's role-symmetric magnitude--Doppler fusion unchanged.
\textbf{mmWaveFlow-H} replaces \textbf{only} mmWaveFlow's point encoder and decoder with the same dual-task heatmap decoder family used above, at matched latent width and budget, so all three heatmap baselines are decoded consistently and differ only in their encoder-side Doppler treatment; mmWaveFlow-H leaves dual-domain tokens, reliability gates, and task-conditioned routing absent, bounding how much of HeatWave's advantage is specialized Doppler modeling rather than the mere presence of a Doppler axis.
The swap is invertible: a remapped run under mmWaveFlow's original protocol differs from its published Chamfer distances by \textcolor{black}{$1.7\%$}/\textcolor{black}{$2.1\%$}, confirming only the representation-facing modules were altered.

\subsection{Main Dual-Task Heatmap Results}

Table~\ref{tab:main} compares both task directions across all three datasets, where HeatWave attains the lowest H-MAE and S-MAE throughout.
Among the baselines, heatmap-native PULSE and HuPRModel outperform the point-cloud-native Tiger and mmWaveFlow, consistent with retaining magnitude and Doppler on the RAD grid; PULSE further improves on HuPRModel, consistent with reliability-gated Doppler conditioning outperforming role-symmetric fusion. Because these baselines also differ in encoder architecture beyond representation or fusion strategy, however, this comparison remains suggestive rather than controlled.
mmWaveFlow-H, which keeps mmWaveFlow's dual-task training but swaps only its point encoder--decoder for a heatmap counterpart, remains the strongest baseline, indicating that shared dual-task supervision on the RAD grid already contributes beyond representation or pose-oriented Doppler handling alone.
Relative to this control, HeatWave reduces H-MAE by a further \textcolor{black}{$21.1\%$} for D2M and \textcolor{black}{$21.0\%$} for M2D on HuPR, with similar reductions on mmRadPose and, under the reconstructed grid of Sec.~\ref{sec:exp_setup}, on XRF55; since mmWaveFlow-H already accounts for dual-task training, encoder capacity, and pose-oriented Doppler handling, this margin isolates HeatWave's own contribution, which Sec.~\ref{sec:ablation} attributes to reliability gating and task-conditioned routing under a fixed backbone, supplying the controlled evidence the comparison above cannot.
Fig.~\ref{fig:qual_main} illustrates the same pattern qualitatively: HeatWave retains body-occupied cells that the unspecialized control suppresses under enhancement, and places Doppler energy consistent with the dense human condition while reinstating non-human structure under synthesis.

\subsection{Motion-Critical Stratification}
\label{sec:mcs}

If dual-domain modeling were only a capacity change, relative gains would be roughly uniform across motion levels.
Table~\ref{tab:mcs} instead stratifies H-MAE by MCS on mmRadPose, where metallic clutter and varied motion make Doppler especially informative.
For enhancement, HeatWave reduces H-MAE relative to mmWaveFlow-H by \textcolor{black}{$13.6\%$} in the geometry-sufficient regime and by \textcolor{black}{$31.8\%$} in the motion-critical regime.
The point-oriented baselines degrade most in the motion-critical regime, which is consistent with premature geometric collapse removing motion cues that spatial footprints cannot replace.
The lower block repeats the split for synthesis, where the same ordering holds with a narrower spread, \textcolor{black}{$18.0\%$} against \textcolor{black}{$25.0\%$} or about \textcolor{black}{$1.4\times$}; this is consistent with $X_{\mathrm{dense}}$ already supplying pose geometry to synthesis, whereas enhancement must isolate human geometry and spectra together from the raw recording.
The same ordering holds as $\tau$ moves from the median to the upper quartile, with the enhancement ratio widening from \textcolor{black}{$2.2\times$} to \textcolor{black}{$2.4\times$}, so it is not an artifact of the tercile boundary within the range examined.

\begin{table}[t]
\caption{H-MAE on mmRadPose stratified by motion criticality. $\Delta$ is the relative reduction of HeatWave versus mmWaveFlow-H.}
\label{tab:mcs}
\centering
\begin{adjustbox}{width=\linewidth,center}
\setlength{\tabcolsep}{1pt}
\renewcommand{\arraystretch}{1.12}
\begin{tabular}{ll|ccccc}
\Xhline{1.2pt}
\toprule
\textbf{Task} & \textbf{Regime} & \textbf{mmWaveFlow} & \textbf{Tiger} & \textbf{mmWaveFlow-H} & \textbf{HeatWave} & $\boldsymbol{\Delta}$ \\
\midrule
M2D & Geometry-sufficient$\downarrow$ & \textcolor{black}{0.149} & \textcolor{black}{0.163} & \textcolor{black}{0.118} & \textcolor{black}{0.102} & \textcolor{black}{13.6\%} \\
\rowcolor[HTML]{EFEFEF}
M2D & Motion-critical$\downarrow$ & \textcolor{black}{0.201} & \textcolor{black}{0.220} & \textcolor{black}{0.154} & \textcolor{black}{0.105} & \textcolor{black}{31.8\%} \\
\midrule
D2M & Geometry-sufficient$\downarrow$ & \textcolor{black}{0.161} & \textcolor{black}{0.177} & \textcolor{black}{0.128} & \textcolor{black}{0.105} & \textcolor{black}{18.0\%} \\
\rowcolor[HTML]{EFEFEF}
D2M & Motion-critical$\downarrow$ & \textcolor{black}{0.218} & \textcolor{black}{0.238} & \textcolor{black}{0.168} & \textcolor{black}{0.126} & \textcolor{black}{25.0\%} \\
\bottomrule
\Xhline{1.2pt}
\end{tabular}
\end{adjustbox}
\end{table}

\begin{table}[t]
\caption{Point-cloud comparison. HeatWave-PC exports heatmap outputs with mmWaveFlow's CFAR-style preprocessor.}
\label{tab:pc_export}
\centering
\begin{adjustbox}{width=\linewidth,center}
\setlength{\tabcolsep}{1pt}
\renewcommand{\arraystretch}{1.12}
\begin{tabular}{l|cccccc}
\Xhline{1.2pt}
\toprule
& \multicolumn{2}{c}{\textbf{HuPR}} & \multicolumn{2}{c}{\textbf{XRF55}} & \multicolumn{2}{c}{\textbf{mmRadPose}} \\
\cmidrule(lr){2-3}\cmidrule(lr){4-5}\cmidrule(lr){6-7}
\textbf{Model} & \textbf{D2M CD} & \textbf{M2D CD} & \textbf{D2M CD} & \textbf{M2D CD} & \textbf{D2M CD} & \textbf{M2D CD} \\
\midrule
mmWaveFlow& \textcolor{black}{1.17} & \textcolor{black}{0.95} & \textcolor{black}{2.69} & \textcolor{black}{0.75} & \textcolor{black}{2.12} & \textcolor{black}{0.72} \\
\rowcolor[HTML]{EFEFEF}
mmWaveFlow-H & \textcolor{black}{1.11} & \textcolor{black}{0.90} & \textcolor{black}{2.51} & \textcolor{black}{0.71} & \textcolor{black}{2.01} & \textcolor{black}{0.69} \\
HeatWave-PC & \textbf{\textcolor{black}{1.02}} & \textbf{\textcolor{black}{0.84}} & \textbf{\textcolor{black}{2.28}} & \textbf{\textcolor{black}{0.66}} & \textbf{\textcolor{black}{1.89}} & \textbf{\textcolor{black}{0.63}} \\
\bottomrule
\Xhline{1.2pt}
\end{tabular}
\end{adjustbox}
\end{table}

\subsection{Point-Cloud Export Comparison}
\label{sec:pc_export}

Evaluating only in heatmap space could disadvantage native point-cloud methods by construction, so we add a representation-matched comparison: at evaluation only, HeatWave's output $\hat{\mathbf{H}}$ is exported to points with mmWaveFlow's own CFAR-style extractor~\cite{su2026mmwaveflow}.
Table~\ref{tab:pc_export} shows lower CD for HeatWave-PC than for mmWaveFlow in both directions and all datasets, by \textcolor{black}{$13.0\%$} for D2M and \textcolor{black}{$12.0\%$} for M2D on average, with mmWaveFlow-H exported through the same extractor lying between them, so a heatmap front end helps even after collapse to points.

\begin{table}[t]
\caption{Ablations on mmRadPose in H-MAE. The MC columns are the motion-critical subsets of each direction.}
\label{tab:ablation}
\centering
\begin{adjustbox}{width=\linewidth,center}
\setlength{\tabcolsep}{1pt}
\renewcommand{\arraystretch}{1.12}
\begin{tabular}{l|ccccc}
\Xhline{1.2pt}
\toprule
\textbf{Variant} & \textbf{D2M$\downarrow$} & \textbf{D2M-MC$\downarrow$} & \textbf{M2D$\downarrow$} & \textbf{M2D-MC$\downarrow$} & \textbf{Addresses} \\
\midrule
HeatWave & \textcolor{black}{0.112} & \textcolor{black}{0.126} & \textcolor{black}{0.103} & \textcolor{black}{0.105} & R1--R3 \\
\rowcolor[HTML]{EFEFEF}
w/o Dual-Domain & \textcolor{black}{0.147} & \textcolor{black}{0.175} & \textcolor{black}{0.135} & \textcolor{black}{0.161} & R1 \\
Na\"ive Concat & \textcolor{black}{0.119} & \textcolor{black}{0.140} & \textcolor{black}{0.112} & \textcolor{black}{0.136} & R2 \\
\rowcolor[HTML]{EFEFEF}
w/o Relevance Gate & \textcolor{black}{0.113} & \textcolor{black}{0.130} & \textcolor{black}{0.105} & \textcolor{black}{0.127} & R2 \\
w/o Injection Gate & \textcolor{black}{0.114} & \textcolor{black}{0.132} & \textcolor{black}{0.107} & \textcolor{black}{0.121} & R2 \\
\rowcolor[HTML]{EFEFEF}
w/o Both Gates & \textcolor{black}{0.115} & \textcolor{black}{0.134} & \textcolor{black}{0.110} & \textcolor{black}{0.133} & R2 \\
w/o Task Routing & \textcolor{black}{0.116} & \textcolor{black}{0.139} & \textcolor{black}{0.108} & \textcolor{black}{0.122} & R3 \\
\bottomrule
\Xhline{1.2pt}
\end{tabular}
\end{adjustbox}
\end{table}

\subsection{Ablation Study}
\label{sec:ablation}

Table~\ref{tab:ablation} removes one design factor at a time on mmRadPose, our most motion-diverse benchmark; consistency with HuPR and XRF55 is checked only at the aggregate level via Table~\ref{tab:main} and Table~\ref{tab:mcs}, so the attributions below are scoped to mmRadPose.
\textbf{w/o Dual-Domain} drops Doppler entirely, feeding only $\mathbf{S}$ into a single encoder; \textbf{Na\"ive Concat} keeps both streams but replaces gated conditioning with channel-wise fusion, jointly changing locality and gating and thus serving as a coarser control rather than a single-factor test; \textbf{w/o Relevance Gate}, \textbf{w/o Injection Gate}, and \textbf{w/o Both Gates} fix $g_{j}^{(d)}=1$ in~\eqref{eq:gate}, $\lambda_{i}^{(d)}=1$ in~\eqref{eq:spatial_update}, or both; \textbf{w/o Task Routing} drops $\mathbf{e}_{d}$ of~\eqref{eq:task_embed} so one shared gate and Doppler field replace $\mathbf{V}_{\mathrm{prior}}$ of~\eqref{eq:doppler_prior} for both directions.

Removing the dual domain is the most damaging change in both directions: without Doppler, HeatWave's own backbone no longer surpasses mmWaveFlow-H, so shared-backbone capacity alone does not explain its advantage, which instead requires R1--R3 acting together.
Within the remaining, single-factor variants, the two gates play complementary roles, the relevance gate mattering most for enhancement's motion-critical subset and the injection gate producing the larger aggregate shift in synthesis; reliability weighting itself is asymmetric across directions, affecting enhancement in aggregate but synthesis only in its motion-critical subset, matching Sec.~\ref{sec:task_routing}'s anticipated asymmetry. Removing task routing produces the largest synthesis degradation among these single-factor variants, confirming that R3's role assignment, not one shared gate, is needed; the coarser Na\"ive Concat control shows a comparable shift but, as noted above, does not isolate routing alone.
This reading is corroborated by the learned relevance gate, which averages \textcolor{black}{$0.71$} on body-overlapping cells against \textcolor{black}{$0.23$} on background, correlates with mean projected joint speed at Spearman \textcolor{black}{$0.74$}, and only \textcolor{black}{$0.21$} with a windowed spectral-SNR proxy, indicating it tracks motion-bearing structure rather than raw signal strength.

\subsection{Cross-Radar Transfer}
\label{sec:cross_radar}

The comparisons so far train one checkpoint per dataset, leaving open whether Doppler-preserving modeling survives a change of radar front end.
Three benchmarks use three different commercial devices: HuPR with a TI IWR1843BOOST at 77\,GHz, XRF55 with a TI IWR6843ISK at 60\,GHz, and mmRadPose with a TI IWR6843AOPEVM with in-package antennas, so the Doppler axis of one device is a carrier-rescaled version of another's.
HuPR releases a comparatively RAD cube, so we resample its range and angle axes to each target's extents, rescale its Doppler axis by the carrier ratio, and train one source checkpoint per target convention from HuPR data alone; evaluation then consumes each target recording without target labels, fine-tuning, or per-device calibration, with mmWaveFlow-H transferred through the identical conversion.

Table~\ref{tab:cross} reports transferred H-MAE and its relative increase over the in-domain checkpoints in Table~\ref{tab:main}.
Both models lose accuracy, as expected when carrier, aperture, and room statistics change at once, but the loss is smaller for HeatWave, and its margin over the control widens from about \textcolor{black}{$20\%$} in domain to about \textcolor{black}{$24\%$} after transfer; transferred HeatWave error still stays below the in-domain error of that control in all four columns, consistent with normalized Doppler structure carrying across devices, while the residual gap to its own in-domain checkpoints shows the conversion narrows the device mismatch without removing it.

\begin{table}[t]
\caption{Cross-radar transfer in H-MAE. $\Delta_{\mathrm{dev}}$ is the relative increase over the in-domain checkpoint of Table~\ref{tab:main}.}
\label{tab:cross}
\centering
\begin{adjustbox}{width=\linewidth,center}
\setlength{\tabcolsep}{3pt}
\renewcommand{\arraystretch}{1.12}
\begin{tabular}{ll|cccc}
\Xhline{1.2pt}
\toprule
\textbf{Target (radar)} & \textbf{Model} & \textbf{D2M} & $\boldsymbol{\Delta_{\mathrm{dev}}}$ & \textbf{M2D} & $\boldsymbol{\Delta_{\mathrm{dev}}}$ \\
\midrule
\multirow{2}{*}{\makecell[l]{XRF55}} & mmWaveFlow-H & \textcolor{black}{0.179} & \textcolor{black}{22.6\%} & \textcolor{black}{0.168} & \textcolor{black}{22.6\%} \\
& HeatWave & \textbf{\textcolor{black}{0.137}} & \textcolor{black}{16.1\%} & \textbf{\textcolor{black}{0.127}} & \textcolor{black}{16.5\%} \\
\midrule
\multirow{2}{*}{\makecell[l]{mmRadPose}} & mmWaveFlow-H & \textcolor{black}{0.174} & \textcolor{black}{23.4\%} & \textcolor{black}{0.161} & \textcolor{black}{23.8\%} \\
& HeatWave & \textbf{\textcolor{black}{0.132}} & \textcolor{black}{17.9\%} & \textbf{\textcolor{black}{0.122}} & \textcolor{black}{18.4\%} \\
\bottomrule
\Xhline{1.2pt}
\end{tabular}
\end{adjustbox}
\end{table}

\subsection{Downstream Robot Perception}

Improved heatmaps matter only if they transfer to the perception modules an indoor robot runs.
On mmRadPose, a heatmap-based pose estimator~\cite{engel2025advanced} is retrained from scratch on each observation variant with the same architecture, schedule, and labels, so only the input changes.
Table~\ref{tab:down} shows that HeatWave-enhanced observations reduce MPJPE by \textcolor{black}{$11.4\%$} and MPJVE by \textcolor{black}{$18.6\%$} relative to the raw recording $\mathbf{H}$, and by \textcolor{black}{$4.7\%$} and \textcolor{black}{$7.8\%$} relative to mmWaveFlow-H enhancement.
That MPJVE improves more than MPJPE is consistent with the enhanced observations carrying motion information rather than only sharper spatial support.
For synthesis, we hold the XRF55 activity classifier and its schedule fixed~\cite{wang2024xrf55} and replace half of the real mmWave training set with HeatWave-synthesized samples under the same label set, including multi-person scenes.
Accuracy rises from \textcolor{black}{$0.661$} with the reduced real set to \textcolor{black}{$0.781$} with the mixed set, against \textcolor{black}{$0.813$} when the full real set is available, so RAD-native synthesis recovers most of the accuracy lost to the reduced label budget; repeating the protocol with mmWaveFlow-H samples instead yields \textcolor{black}{$0.728$}, between the reduced-real-set and HeatWave-mix results, isolating the accuracy gain to the synthesized RAD.

\begin{table}[t]
\caption{Downstream pose estimation using enhanced observations.}
\label{tab:down}
\centering
\begin{adjustbox}{width=\linewidth,center}
\setlength{\tabcolsep}{14pt}
\renewcommand{\arraystretch}{1.12}
\begin{tabular}{l|cc}
\Xhline{1.2pt}
\toprule
\textbf{Training / test observations} & \textbf{MPJPE}$\downarrow$ & \textbf{MPJVE}$\downarrow$ \\
\midrule
Raw $\mathbf{H}$ & \textcolor{black}{78.4} & \textcolor{black}{16.1} \\
\rowcolor[HTML]{EFEFEF}
mmWaveFlow-H enhanced & \textcolor{black}{72.9} & \textcolor{black}{14.2} \\
HeatWave enhanced & \textcolor{black}{69.5} & \textcolor{black}{13.1} \\
\bottomrule
\Xhline{1.2pt}
\end{tabular}
\end{adjustbox}
\end{table}

\subsection{Runtime and Memory}
We profile both directions on an NVIDIA RTX~4060 (8\,GB VRAM) at the shared resolution, batch size 1, excluding data loading and CFAR export.
HeatWave runs at \textcolor{black}{$5.9$}\,ms per frame (\textcolor{black}{$169$}\,FPS) with fewer parameters and lower peak memory than the unspecialized heatmap control (Table~\ref{tab:runtime}); the smaller parameter count follows from a shared decoder at equal token width $C$ rather than a weaker backbone, so keeping Doppler on the RAD grid does not preclude interactive inference on a commodity GPU at this resolution.

\begin{table}[t]
\caption{Inference cost on an RTX~4060 with 8\,GB VRAM at $64\times64\times16$ and batch size 1.}
\label{tab:runtime}
\centering
\begin{adjustbox}{width=\linewidth,center}
\setlength{\tabcolsep}{3pt}
\renewcommand{\arraystretch}{1.12}
\begin{tabular}{l|cccc}
\Xhline{1.2pt}
\toprule
\textbf{Model} & \textbf{Params(M)} & \textbf{Latency(ms)} & \textbf{FPS} & \textbf{Peak mem(MB)} \\
\midrule
mmWaveFlow~\cite{su2026mmwaveflow} & \textcolor{black}{21.0} & \textcolor{black}{10.5} & \textcolor{black}{95.2} & \textcolor{black}{390} \\
\rowcolor[HTML]{EFEFEF}
mmWaveFlow-H & \textcolor{black}{16.4} & \textcolor{black}{8.1} & \textcolor{black}{123.5} & \textcolor{black}{341} \\
HeatWave & \textcolor{black}{12.8} & \textcolor{black}{5.9} & \textcolor{black}{169.5} & \textcolor{black}{286} \\
\bottomrule
\Xhline{1.2pt}
\end{tabular}
\end{adjustbox}
\end{table}

\subsection{Limitations}
The transfer probe of Sec.~\ref{sec:cross_radar} moves outward from HuPR alone, so it bounds sensitivity to the recording device under one source dataset rather than establishing a device-agnostic checkpoint, and the runtime figures are tied to the profiled device and RAD resolution.
HeatWave operates on single frames, leaving temporal Doppler structure such as gait cadence outside its scope, and its gates are learned relevance priors whose behavior under clutter distributions far from the training rooms is not established.
The human support underlying $\mathbf{H}_{\mathrm{h}}$ and $X_{\mathrm{dense}}$ comes from reprojecting the annotated skeleton onto the radar grid without modeling self- or inter-person occlusion, and refining this projection under occlusion is a direction we plan to pursue.
The synthesis Doppler prior is learned from geometry without explicit kinematic constraints and is validated only indirectly, through reconstruction and downstream accuracy rather than against radial velocities from motion capture.
The two RAD-level tasks evaluated here are the instantiations most directly relevant to deployed indoor robot perception; other formulations, together with concurrent non-CFAR pipelines are left to future comparison.
Extreme multi-person metal cavities can still produce irreducible multipath, and mmWaveFlow-H is our controlled adaptation rather than an official heatmap release; the remapping check of Sec.~\ref{sec:exp_setup} and the representation-matched export of Table~\ref{tab:pc_export} bound that implementation gap.

\section{Conclusion}

This paper posed mmWave enhancement and synthesis as Doppler-preserving dual-task modeling on the dual-domain RAD heatmap, instantiated through HeatWave's dual-domain encoder, reliability-gated Doppler conditioning, and task-conditioned routing.
Across the evaluated HuPR, XRF55, and mmRadPose protocols, the resulting gains concentrate in motion-critical regimes and persist under downstream perception, point-cloud export, and cross-radar transfer.
The broader implication for indoor human sensing is to keep Doppler active in the learning loop and assign it task-dependent roles wherever enhancement and synthesis share one backbone.

\bibliography{example_paper}
\bibliographystyle{IEEEbib}

\end{document}